\documentclass[aps,prd,twocolumn,superscriptaddress,nofootinbib,floatfix]{revtex4-2}

\usepackage{amsmath,amssymb}
\usepackage{graphicx}
\usepackage{bm}
\usepackage{xcolor}
\usepackage[colorlinks=true,allcolors=blue!55!black]{hyperref}
\usepackage{orcidlink}

\newcommand{\Qd}{\dot{Q}}
\newcommand{\wtot}{w_{\rm tot}}
\newcommand{\rhoeff}{\rho_{\rm eff}}
\newcommand{\peff}{p_{\rm eff}}

\begin{document}

\title{Viscosity heats, diffusion drains: Open-system thermodynamics in unimodular cosmology}

\author{Miguel Cruz\orcidlink{0000-0003-3826-1321}}
\email{miguelcruz02@uv.mx}
\affiliation{Facultad de F\'isica, Universidad Veracruzana 91097, Xalapa, Veracruz, M\'exico}

\author{Samuel Lepe\orcidlink{0000-0002-3464-8337}}
\email{samuel.lepe@pucv.cl}
\affiliation{Instituto de F\'isica, Pontificia Universidad Cat\'olica de Valpara\'iso,
Casilla 4950, Valpara\'iso, Chile}

\date{\today}

\begin{abstract}
In general relativity with bulk viscosity and in unimodular gravity with energy diffusion, the matter sector is non-adiabatic, and its balance, entropy and temperature equations coincide under the identification $\Pi\leftrightarrow\dot Q/3H$. Whether a unimodular cosmology is adiabatic has nevertheless been answered in opposite ways, depending on whether entropy is assigned to matter alone or to the total effective fluid. We show that the answer is fixed by the temperature $T_D$ of the sector that receives the diffused energy: the total entropy production per comoving volume is $\dot Q\,(1/T_D-1/T_m)$, recovering the previous proposals in the limits $T_D\to\infty$ and $T_D\to T_m$. Adopting the microscopic picture in which unimodular dissipation feeds a bath of hidden degrees of freedom at finite $T_D$, diffusion acts as heat conduction to an external reservoir: whereas bulk viscosity can only heat a closed fluid, diffusion into a cold bath cools an open one. This structural difference breaks the correspondence with viscosity at the level of observables and stability. Cosmographically, the viscous pressure is fixed by the deceleration parameter $q$ and the diffusion rate by the jerk $j$, making $\Lambda$CDM the adiabatic limit of unimodular cosmology but permanently dissipative as a viscous fluid. Moreover, the diffusive analogue of the near-equilibrium ratio $|\Pi|/\rho$ is the shift $c_s^2-\omega$ of the effective sound speed, which is governed by perturbative stability rather than cosmic acceleration. When dark matter is modeled as a gas with conserved rest mass, thermodynamically consistent diffusion removes thermal energy alone, precludes phantom behavior, and bounds the growth of the cosmological term by the available thermal reservoir; consequently, realizing the logotropic dark fluid with conserved particle number would require relativistic dark matter at $z\simeq12$. Finally, using the dynamical Hayward--Kodama temperature of the apparent horizon, the generalized second law follows directly from the local one in unimodular gravity, whereas for bulk viscosity it remains an independent constraint.
\end{abstract}

\maketitle

%=====================================================================
\section{Introduction}
\label{sec:intro}
%=====================================================================
Unimodular gravity (UG) is the simplest modification of general relativity (GR) that can
accommodate dissipative effects at the level of the field equations
\cite{Unruh:1989,Ellis:2011,Carballo-Rubio:2022,Bengochea:2023}. Fixing the metric
determinant replaces the Einstein equations by their trace-free part, whose Bianchi identity no
longer enforces $\nabla^{\nu}T_{\mu\nu}=0$; instead, a diffusion current
$J_\mu=\nabla^{\nu}T_{\mu\nu}$ sources a cosmological term $\Lambda=\Lambda_0+\int_\ell J$, and the
limit $J=0$ is GR with a cosmological constant \cite{Josset:2017,Perez:2019}. The resulting
cosmologies have been studied with a variety of phenomenological diffusion laws
\cite{Garcia-Aspeitia:2019,Corral:2020,Perez:2021,Linares:2021,Landau:2023,Cruz:2024}, and
recently with a law derived from a microscopic model in which matter dissipates into a bath of
hidden, Planckian degrees of freedom \cite{Pellecchia:2026}.

On the GR side, the canonical way of introducing dissipation in a homogeneous background is a
bulk viscous pressure $\Pi$, governed by the Eckart \cite{Eckart:1940} or the causal
Israel--Stewart \cite{Israel:1979,Hiscock:1985} theory
\cite{Maartens:1995,Maartens:1996,Brevik:2017}. The two approaches share several features: each
renders the matter sector non-adiabatic, each can drive accelerated expansion with pressureless
matter alone, and in both the balance law, the entropy budget of matter and the temperature
equation are mapped into one another by $\Pi\leftrightarrow\dot Q/3H$
(Sec.~\ref{sec:equations}). This raises the question of whether the match is more than a formal
analogy.

The issue is more subtle than it first appears, since the thermodynamic status of unimodular diffusion remains unresolved. Writing the Gibbs relation for the conserved total effective fluid,
Ref.~\cite{Cruz:2026logo} concluded that the unimodular universe is adiabatic, so that a
diffusion rate of either sign is compatible with the second law. Writing it for the matter
sector alone, one finds the entropy production $nT\dot\sigma=-\dot Q$ \cite{Cardenas:2026} and,
from the requirement that this entropy not decrease, one can constraint the diffusion function
\cite{Cruz:2024}. Both computations are correct, but they refer to different systems. The
apparent conflict disappears once one specifies what entropy, if any, is carried by the sector
that receives the diffused energy, and the same specification decides whether
$\Pi\leftrightarrow\dot Q/3H$ holds beyond the matter balance equations.

Our answer is that the receiving sector must have a temperature $T_D$. The total entropy
production is then $\dot Q\,(1/T_D-1/T_m)$ per comoving volume, and the two previous approaches are its limits $T_D\to\infty$ and $T_D\to T_m$ (Sec.~\ref{sec:accountings}). The microscopic model of Ref.~\cite{Pellecchia:2026} provides precisely such a sector, a bath obeying a fluctuation-dissipation relation, and we adopt it as the physical accounting. Diffusion then becomes heat conduction between matter and an external reservoir, turning the formal match with internal bulk viscosity into a physical anti-correspondence (Sec.~\ref{sec:heats}). As summarized in Table~\ref{tab:dictionary}, we explore the resulting physical differences across four domains: the cosmographic reconstruction and effective sound speed of the system (Sec.~\ref{sec:cosmography}), the thermal reservoir bound when dark matter is modeled as a gas with conserved rest mass and its implications for phantom behavior and the logotropic dark fluid (Sec.~\ref{sec:positivity}), and the generalized second law at the dynamical apparent horizon (Sec.~\ref{sec:horizon}). We discuss open problems in Sec.~\ref{sec:discussion} and synthesize our conclusions in Sec.~\ref{sec:conclusions}.

We work with a spatially flat Friedmann-Lema\^itre-Robertson-Walker (FLRW) metric (see
Ref.~\cite{Aguilar:2026} for curvature in UG), units $8\pi G=c=\hbar=k_B=1$, and a barotropic matter sector $p=\omega\rho$ with constant parameter of state $\omega>-1$ and $\rho>0$ initially, except in Sec.~\ref{sec:positivity}, where dark matter is treated as a gas with rest mass. The overdots denote cosmic time derivatives, $H=\dot a/a$, and the deceleration and jerk parameters are $q=-1-\dot H/H^2$ and $j=\dddot{a}/(aH^3)$, so that
\begin{equation}
\dot H=-(1+q)H^2,\qquad \ddot H=(j+3q+2)H^3.
\label{eq:kinematics}
\end{equation}
A subscript $0$ denotes present-time values, and we normalize $a_0=1$. In UG we set
$\Lambda_0=0$, which merely shifts $Q$ by a constant.

%=====================================================================
\section{Field equations and the formal correspondence}
\label{sec:equations}
%=====================================================================
In GR with a bulk viscous pressure $\Pi$,
\begin{gather}
3H^2=\rho,\qquad \dot H=-\tfrac12\left(\rho+p+\Pi\right),\label{eq:GR1}\\
\dot\rho+3H(\rho+p)=-3H\Pi,\label{eq:GR2}
\end{gather}
and in UG with energy diffusion,
\begin{gather}
3H^2=\rho+Q,\qquad \dot H=-\tfrac12\left(\rho+p\right),\label{eq:UG1}\\
\dot\rho+3H(\rho+p)=-\Qd,\label{eq:UG2}
\end{gather}
where $Q$ is the dynamical part of the cosmological term. Inserting the Raychaudhuri equations
into $q=-1-\dot H/H^2$ gives the deceleration parameters
$q_{\rm GR}=\tfrac12[1+3(\omega+\Pi/\rho)]$ and $q_{\rm UG}=(\rho+3p-2Q)/[2(\rho+Q)]$. The
structural asymmetry is already visible: $Q$ enters the constraint and not the Raychaudhuri
equation, whereas $\Pi$ enters the Raychaudhuri equation and not the constraint. Only $\Qd$ is directly comparable to $\Pi$, because a constant $Q$ corresponds to a cosmological constant. In terms of the redshift $z=a^{-1}-1$, Eqs.~\eqref{eq:GR2} and \eqref{eq:UG2} integrate from $z=0$ to the common form
\begin{equation}
\rho(z)=\rho_0(1+z)^{3(1+\omega)}\left[1+\frac{1}{\rho_0}\int_0^z
\frac{s(z')\,dz'}{(1+z')^{3(1+\omega)}}\right],
\label{eq:rhoz}
\end{equation}
with $s=3\Pi/(1+z)$ in GR and $s=-dQ/dz$ in UG. Equivalently, the balance laws describe the dilution of a fluid with a modified parameter of state,
\begin{equation}
\dot\rho+3H(1+\bar\omega)\rho=0,\qquad \bar\omega=\omega+\frac{\Pi}{\rho},
\label{eq:wbar}
\end{equation}
\begin{equation}
\dot\rho+3H(1+\tilde\omega)\rho=0,\qquad \tilde\omega=\omega+\frac{\Qd}{3H\rho}.
\label{eq:wtilde}
\end{equation}

The thermodynamics of the matter sector follows from the Gibbs relation per particle,
$T_m\,d\sigma=d(\rho/n)+p\,d(1/n)$, where $\sigma$ is the entropy per particle and $T_m$ the
matter temperature. With conserved particle number, $\dot n=-3Hn$, and fixed particle mass, it
gives $nT_m\dot\sigma=\dot\rho-(\rho+p)\dot n/n=\dot\rho+3H(\rho+p)$. The entropy
$S_m=nV\sigma$ in a comoving volume $V\propto a^3$ then obeys, by Eqs.~\eqref{eq:GR2} and
\eqref{eq:UG2},
\begin{equation}
\frac{T_m\dot S_m}{V}=-3H\Pi\ \ \text{(GR)},\qquad
\frac{T_m\dot S_m}{V}=-\Qd\ \ \text{(UG)}.
\label{eq:Sm}
\end{equation}
The integrability of the Gibbs relation fixes the evolution of the matter temperature
\cite{Maartens:1996} (Appendix~\ref{app:temperature}),
\begin{equation}
\frac{\dot T_m}{T_m}=-3H\left(\frac{\partial p}{\partial\rho}\right)_{\!n}
-\frac{\mathcal{F}}{T_m}\left(\frac{\partial T_m}{\partial\rho}\right)_{\!n},
\label{eq:Tdot}
\end{equation}
with $\mathcal F=3H\Pi$ in GR and $\mathcal F=\Qd$ in UG. Equations
\eqref{eq:rhoz}--\eqref{eq:Tdot} are all interchanged by
\begin{equation}
\Pi\;\longleftrightarrow\;\frac{\Qd}{3H}.
\label{eq:map}
\end{equation}

These equations exhaust the correspondence: it fails as soon as the Friedmann constraint is
used. Two examples show how. First, $\bar\omega$ is the total parameter of state of the viscous
universe, since $3H^2=\rho$, whereas $\tilde\omega$ is not the total parameter of state of UG but
the adiabatic sound speed of its effective fluid (Sec.~\ref{sec:soundspeed}). Second, suppose
that the expansion history $H(t)$ of a unimodular model is described in GR by matter with the
same $\omega$ plus bulk viscosity. The constraint~\eqref{eq:GR1} assigns to this matter the
density $3H^2=\rho+Q$, and the Raychaudhuri equations~\eqref{eq:GR1} and \eqref{eq:UG1} then
require $(1+\omega)(\rho+Q)+\Pi=(1+\omega)\rho$. The viscous pressure that reproduces the
unimodular history is therefore
\begin{equation}
\Pi=-(1+\omega)\,Q,
\label{eq:Piinf}
\end{equation}
set by $Q$ itself and not by $\Qd/3H$.

For later use we recall the effective-fluid form of UG \cite{Cruz:2026logo}. With
$\rhoeff=\rho+Q$ and $\peff=p-Q$, Eqs.~\eqref{eq:UG1} read $3H^2=\rhoeff$ and
$\dot H=-\tfrac12(\rhoeff+\peff)$, and adding $\Qd$ to both sides of Eq.~\eqref{eq:UG2} gives
\begin{equation}
\dot\rho_{\rm eff}+3H(\rhoeff+\peff)=0
\label{eq:effcons}
\end{equation}
identically. The total parameter of state is $\wtot=\peff/\rhoeff=(\omega\rho-Q)/(\rho+Q)$, with
\begin{equation}
1+\wtot=\frac{(1+\omega)\rho}{\rho+Q}.
\label{eq:wtot}
\end{equation}
In this language, Eq.~\eqref{eq:Piinf} reads $\Pi=\peff-\omega\rhoeff$: the viscous pressure is
the part of the effective pressure not accounted for by the parameter of state of matter.

%=====================================================================
\section{Open-system thermodynamics and the three entropy accountings}
\label{sec:accountings}
%=====================================================================
\subsection{Open-system balance and the three thermodynamic regimes}
\label{sec:open}

The unimodular matter sector is a perfect fluid: its non-adiabaticity comes from the balance law
$\nabla^\nu T_{\mu\nu}=J_\mu$, not from a correction to its stress tensor. To determine the total entropy budget in a comoving volume $V\propto a^3$ (with $dV=3HV\,dt$) during an interval $dt$ in which $dQ=\Qd\,dt$, we track the energy and entropy across sectors. For the matter contained in $V$, the Gibbs relation $T_m\,dS_m=d(\rho V)+p\,dV$ combined with Eq.~\eqref{eq:UG2} gives
\begin{equation}
T_m\,dS_m=V\left[\dot\rho+3H(\rho+p)\right]dt=-V\,dQ,
\label{eq:Smatter}
\end{equation}
which is Eq.~\eqref{eq:Sm} in differential form: for $dQ>0$, matter releases the heat $V\,dQ$ independently of any microscopic interpretation of $Q$. Meanwhile, the geometric sector behaves macroscopically as a fluid with $\rho_Q=Q$ and $p_Q=-Q$, whose first-law combination is
\begin{equation}
d(\rho_QV)+p_Q\,dV=V\,dQ+Q\,dV-Q\,dV=V\,dQ .
\label{eq:Qenergy}
\end{equation}
The sum of $d(\rho V)+p\,dV=-V\,dQ$ and Eq.~\eqref{eq:Qenergy} vanishes, recovering the total conservation law~\eqref{eq:effcons}: the energy released by matter is transferred to the geometric sector.

Equation~\eqref{eq:Qenergy} fixes where the energy goes, but not the entropy that accompanies it. Assigning that entropy to hidden degrees of freedom at temperature $T_D$ that absorb the heat $V\,dQ$ (as shown below in Sec.~\ref{sec:where}, the geometric term $Q$ itself cannot carry entropy), their entropy changes by
\begin{equation}
T_D\,dS_D=V\,dQ .
\label{eq:SD}
\end{equation}
Adding $dS_m=-V\,dQ/T_m$ and $dS_D=V\,dQ/T_D$ yields the total entropy production per unit volume,
\begin{equation}
\frac{\dot S_m+\dot S_D}{V}=\Qd\left(\frac{1}{T_D}-\frac{1}{T_m}\right).
\label{eq:Stot}
\end{equation}

The entropy accountings used in the literature are particular regimes of Eq.~\eqref{eq:Stot}:
\begin{itemize}
\item[(A)] \emph{Matter limit} ($T_D\to\infty$): the receiving sector absorbs heat without changing its entropy, the total entropy is that of matter alone, and the second law requires $\Qd\le0$ \cite{Cruz:2024,Cardenas:2026}.
\item[(B)] \emph{Reversible limit} ($T_D=T_m$): the total entropy is conserved for any $\Qd$, recovering the adiabatic accounting of Ref.~\cite{Cruz:2026logo}.
\item[(C)] \emph{Finite-bath regime} (finite $T_D\neq T_m$): the second law requires
\begin{equation}
\Qd\,(T_m-T_D)\ge0,
\label{eq:secondlawC}
\end{equation}
so that heat flows from the hotter to the colder sector. Its cold-bath limit $T_D\to0$ is the mirror image of the matter limit (A), admitting $\Qd\ge0$ and forbidding $\Qd<0$.
\end{itemize}
Crucially, total energy conservation~\eqref{eq:effcons} does not by itself imply the reversible regime (B). Reference~\cite{Cruz:2026logo} obtained $dS=0$ by writing the Gibbs relation directly for the conserved effective fluid $(\rhoeff,\peff)$. The same argument would render every Eckart viscous cosmology adiabatic, since $\dot\rho+3H(\rho+p+\Pi)=0$ takes the conserved perfect-fluid form with total pressure $p+\Pi$, even though the physical entropy grows according to Eq.~\eqref{eq:Sm}. The Gibbs relation holds only when written in terms of the \emph{equilibrium} pressure of the system, and neither $p+\Pi$ nor $\peff=p-Q$ is an equilibrium pressure unless internal relaxation has ceased or the two sectors share a common temperature $T_D=T_m$. Consequently, while the background dictionary of Ref.~\cite{Cruz:2026logo} between UG and the adiabatic logotropic dark fluid \cite{Chavanis:2016} remains mathematically exact, its adiabatic reading represents a specific hypothesis about the bath temperature, $T_D=T_m$.

\subsection{Microscopic bath and linear-response closure}
\label{sec:where}

Can the geometric sector $\rho_v=Q$, $p_v=-Q$ carry the entropy directly without invoking a hidden bath? Treating it as a fluid at temperature $T_v$ with vanishing chemical potential, the integrability condition $T_v\,\partial p_v/\partial T_v=\rho_v+p_v=0$ forces its entropy density $s_v=(\rho_v+p_v)/T_v$ to vanish ($S_v=0$), whereas its first law~\eqref{eq:Qenergy} requires
\begin{equation}
T_v\dot S_v=\frac{d}{dt}(\rho_vV)+p_v\dot V=V\Qd,
\label{eq:vacuum}
\end{equation}
which is consistent only if $\Qd=0$. Thus, no thermal history can be assigned to $\Lambda+Q(t)$ itself if it retains the equation of state of a cosmological constant, invalidating a naive analogy with an entropy-carrying interacting vacuum. The heat $V\,dQ$ must reside in microscopic degrees of freedom that do not obey $p=-\rho$ internally; in UG, the trace-free field equations and the Bianchi identity integrate the divergence $J_\mu=\nabla^\nu T_{\mu\nu}$ strictly into the effective cosmological term $\Lambda+Q$, which acts as the macroscopic gravitational imprint of the reservoir without coupling the bath's internal thermal pressure to the background Raychaudhuri equation. Reference~\cite{Pellecchia:2026} reached this same picture from the unimodular Hamiltonian. In minisuperspace the cosmological constant is the conserved total Hamiltonian, $\Lambda_0=-8\pi G\mathcal{H}$, so a dynamical $\Lambda$ requires hidden degrees of freedom $\mathcal{H}_D$ coupled to matter. For a bath of oscillators with Ohmic damping \cite{Caldeira:1983,Calzetta:2008}, a matter field $\phi$ obeys a Langevin equation with friction $\bar\beta m_p\dot\phi$ ($m_p$ being the Planck mass) and thermal noise $\xi$ governed by the fluctuation-dissipation theorem at the bath temperature. The diffusion rate takes the schematic form $\Qd\propto\bar\beta m_p\dot\phi^2-\dot\phi\,\xi$, which in the noiseless (cold-bath) limit yields
\begin{equation}
\Qd=\frac{\gamma m_p}{2}\left(\frac{a_d}{a}\right)^{3}(\rho+p)\ \ge0,
\label{eq:PPR}
\end{equation}
where $\gamma$ and $a_d$ are constants. Including thermal noise opposes the dissipative drain so that net exchange ceases when matter equilibrates with the bath (Appendix~A of Ref.~\cite{Pellecchia:2026}), providing a microscopic realization of the finite-bath regime (C). From here on we adopt the finite-bath regime (C) with the minimal linear-response (Newton) closure compatible with Eq.~\eqref{eq:secondlawC},
\begin{equation}
\Qd=K\,(T_m-T_D),\qquad K\ge0,
\label{eq:Newton}
\end{equation}
where $K$ may depend on $T_m$, $T_D$, $n$ and $a$ (for instance, a radiative exchange $\Qd\propto T_m^4-T_D^4$ corresponds to $K\propto(T_m+T_D)(T_m^2+T_D^2)$). Substituting Eq.~\eqref{eq:Newton} into Eq.~\eqref{eq:Stot} gives a quadratic entropy production,
\begin{equation}
\frac{\dot S_m+\dot S_D}{V}=\frac{K\,(T_m-T_D)^2}{T_mT_D}=\frac{\Qd^{\,2}}{K\,T_mT_D}\ \ge0 .
\label{eq:Squad}
\end{equation}
This relation clarifies the status of the reversible limit (B): sustaining a finite flux $\Qd\neq 0$ at $T_m=T_D$ with zero entropy production requires $K\to\infty$. For any finite $K$, thermal equilibrium $T_m=T_D$ implies $\Qd=0$, in full agreement with the fluctuation-dissipation structure of Ref.~\cite{Pellecchia:2026}.

%=====================================================================
\section{Viscosity heats, diffusion drains}
\label{sec:heats}

\begin{table*}[t]
\begin{ruledtabular}
\begin{tabular}{lll}
 & GR, bulk viscosity & UG, diffusion into a bath at $T_D$\\
\hline
System & closed: one imperfect fluid & open: perfect-fluid matter plus hidden bath\\
Enters & Raychaudhuri equation & Friedmann constraint\\
Flux / conjugate force & $\Pi$ / expansion rate $3H$ & $\Qd$ / $1/T_D-1/T_m$\\
Constitutive law & $\Pi=-3\zeta H$; relaxation $\tau$ in Israel--Stewart & $\Qd=K(T_m-T_D)$\\
Entropy production & $9\zeta H^2/T_m\ge0$ & $K(T_m-T_D)^2/(T_mT_D)\ge0$\\
Effect on matter & always heats ($\Pi\le0$) & heats or drains; drains into a cold bath\\
Equilibrium & no expansion, or $\zeta=0$ & $T_m=T_D$\\
\hline
Cosmography & $q$: $\Pi=H^2(2q-1-3\omega)$ & $j$: $\Qd=2H^3[j-1-3\omega(1+q)]/(1+\omega)$\\
Adiabatic limit (dust) & $q=1/2$ & $j=1$ ($\Lambda$CDM)\\
Smallness parameter & $|\Pi|/\rho$; conflicts with acceleration & $\Qd/3H\rho=c_s^2-\omega$; unrelated to acceleration\\
Energy available & work of the expansion, unbounded & heat of matter, $\Delta Q\le\varepsilon$ (not rest mass)\\
Phantom behavior & allowed by local second law; excluded by GSL & excluded by local second law\\
GSL at the apparent horizon & independent constraint ($q\ge -1$ for $q<1$) & implied by the local second law (for $q<1$)\\
\end{tabular}
\end{ruledtabular}
\caption{Bulk viscosity in GR and energy diffusion in UG under the finite-bath regime (C). Rows above
the line are structural; rows below are the consequences derived in
Secs.~\ref{sec:cosmography}--\ref{sec:horizon}. Entropy production is per comoving volume, and
GSL stands for the generalized second law.}
\label{tab:dictionary}
\end{table*}

Both mechanisms produce entropy as the product of a flux and a thermodynamic force, but the
forces are of different kinds. For viscosity, the Israel--Stewart entropy current
$S^a=n\sigma u^a-(\tau\Pi^2/2\zeta T)u^a$ and the condition $\nabla_aS^a\ge0$ yield the transport
equation
\begin{equation}
\tau\dot\Pi+\Pi=-3\zeta H-\frac{\tau\Pi}{2}\left[3H+\frac{\dot\tau}{\tau}-\frac{\dot\zeta}{\zeta}
-\frac{\dot T}{T}\right],
\label{eq:IS}
\end{equation}
with bulk viscosity $\zeta\ge0$ and relaxation time $\tau>0$ \cite{Israel:1979,Maartens:1996}.
The force is the expansion rate, the dissipation is internal to a closed fluid, and in an
expanding universe $\Pi\le0$. By Eq.~\eqref{eq:GR2}, viscosity can only \emph{heat} the matter
sector, $\dot\rho+3H(\rho+p)=-3H\Pi\ge0$. For diffusion in the finite-bath regime (C), the force is the temperature
difference with an external system, Eq.~\eqref{eq:Stot}, and heat flows from the hotter to the
colder sector. For the cold bath of the microscopic model, $\Qd>0$ and diffusion \emph{drains} the
matter sector, $\dot\rho+3H(\rho+p)=-\Qd<0$.

The map~\eqref{eq:map} therefore sends the physical sign of viscosity, $\Pi<0$, to $\Qd<0$, which
in the finite-bath regime (C) requires a bath hotter than matter, and it sends the physical sign of cold-bath diffusion,
$\Qd>0$, to $\Pi>0$, which viscosity forbids. For the microscopically motivated case the
correspondence is an anti-correspondence. In mechanical terms, viscosity is the internal
resistance of a fluid to its own expansion, whereas diffusion is the Brownian friction
$\bar\beta m_p\dot\phi$ against an external reservoir of Ref.~\cite{Pellecchia:2026}, which comes
with its fluctuation-dissipation partner $\xi$.

A further difference is often attributed to the order of the dynamics. In Israel--Stewart theory
$\Pi$ is an independent variable, and causality requires that the propagation speed of viscous signals, $c_b^2=\zeta/[\tau(\rho+p)]$, added to the sound speed of matter, $\omega$, not exceed the speed of light \cite{Maartens:1996},
\begin{equation}
\frac{\zeta}{\tau(\rho+p)}\le1-\omega .
\label{eq:causal}
\end{equation}
For a given history $H(t)$ with $\Pi=-2\dot H-3(1+\omega)H^2<0$ and the truncated transport law $\tau\dot\Pi+\Pi=-3\zeta H$, simultaneously satisfying $\zeta\ge0$ and Eq.~\eqref{eq:causal} requires
\begin{equation}
\tau\dot\Pi\le-\Pi\le\tau\left[\dot\Pi+9(1-\omega^2)H^3\right].
\label{eq:twoconds}
\end{equation}
A causal relaxation time $\tau>0$ therefore exists if and only if $\dot\Pi+9(1-\omega^2)H^3>0$ (leaving the Eckart limit $\tau\to0$ non-causal for $\Pi<0$). Using Eq.~\eqref{eq:kinematics} to evaluate $\dot\Pi=2H^3[3(1+\omega)(1+q)-j-3q-2]$, this existence condition translates into the cosmographic upper bound
\begin{equation}
j<\tfrac12\left(11+6\omega+6\omega q-9\omega^2\right),
\label{eq:jbound}
\end{equation}
i.e., $j<11/2$ for pressureless matter ($\omega=0$). This higher-order constraint, however, belongs to the Israel--Stewart closure rather than to the GR frame itself. The
fair comparison is between Eckart's law $\Pi=-3\zeta H$ and the Newton law~\eqref{eq:Newton}: both are instantaneous closure relations where the flux is determined by the force at the same instant and has no memory effects. We comment on a causal extension of Eq.~\eqref{eq:Newton} in Sec.~\ref{sec:discussion}.

%=====================================================================
\section{Kinematic and perturbative signatures: Cosmography and sound speed}
\label{sec:cosmography}
%=====================================================================
\subsection{Cosmographic reading: Deceleration versus jerk}

Given an expansion history $H(t)$ and the parameter of state $\omega$, each frame infers its dissipative
source from a different derivative of the scale factor. In GR, Eq.~\eqref{eq:GR1} gives $\Pi=-2\dot H-(1+\omega)\rho$ with $\rho=3H^2$, which Eq.~\eqref{eq:kinematics} and Eq.~\eqref{eq:Sm} express in terms of the deceleration parameter $q$:
\begin{equation}
\Pi=H^2\left(2q-1-3\omega\right),\quad
\frac{T_m\dot S_m}{V}=3H^3\left(1+3\omega-2q\right).
\label{eq:Piq}
\end{equation}
In UG, the Raychaudhuri equation~\eqref{eq:UG1} fixes $\rho=-2\dot H/(1+\omega)$ and the
Friedmann constraint gives $Q=3H^2-\rho=[2\dot H+3(1+\omega)H^2]/(1+\omega)$. Differentiating and
using Eq.~\eqref{eq:kinematics} relates the diffusion rate to the jerk $j$:
\begin{equation}
\Qd=\frac{2\ddot H+6(1+\omega)H\dot H}{1+\omega}=\frac{2H^3}{1+\omega}\left[j-1-3\omega(1+q)\right].
\label{eq:Qdj}
\end{equation}
For pressureless dark matter ($\omega=0$),
\begin{equation}
\Pi=H^2(2q-1),\qquad \Qd=2H^3(j-1).
\label{eq:dust}
\end{equation}
Because $\Pi$ enters the Raychaudhuri equation whereas $Q$ enters the Friedmann constraint, the viscous pressure is read from the second derivative of $a(t)$ ($q$) and the diffusion rate from the third ($j$). (For strictly pressureless matter, $\Qd\neq 0$ in Eq.~\eqref{eq:dust} implies a change of rest-mass energy; the generalization for a gas with rest mass and thermal energy is given in Eq.~\eqref{eq:Qdjgas}.)

Three physical differences follow immediately:
\begin{enumerate}
\item \emph{Different adiabatic limits:} By Eq.~\eqref{eq:Piq}, the viscous reading is adiabatic only for a decelerating perfect fluid $q=(1+3\omega)/2$, whereas by Eq.~\eqref{eq:Qdj} the diffusive reading is adiabatic whenever $j=1+3\omega(1+q)$ ($j=1$ for dust). Cosmic acceleration requires dissipation in the viscous frame, but not in UG.
\item \emph{Thermodynamic status of $\Lambda$CDM:} Since $\Lambda$CDM has $j=1$, it represents the adiabatic limit ($Q={\rm const}$) of the unimodular cosmology. However, read as a viscous fluid with $\omega=0$, reproducing $\Lambda$CDM requires $\Pi=-\Lambda$ at all times by Eq.~\eqref{eq:Piinf}; for $q_0\simeq-0.55$, Eq.~\eqref{eq:dust} yields $|\Pi|/\rho=(1-2q_0)/3\simeq0.7$, far outside the regime of quasi-equilibrium, with permanent production of entropy $T_m\dot S_m/V=3H^3(1-2q)>0$.
\item \emph{Thermal meaning of the jerk:} In the finite-bath regime (C) with $K>0$, the sign of Eq.~\eqref{eq:Qdj} tracks the sign of $T_m-T_D$. For nearly pressureless matter, $j>1$ holds if and only if matter is hotter than the bath and is drained by it, up to thermal corrections of order $\varepsilon/\rho$ [Eq.~\eqref{eq:Qdjgas}].
\end{enumerate}

\subsection{The near-equilibrium ratio as an effective sound speed}
\label{sec:soundspeed}

In viscous cosmology, truncating the Israel--Stewart series requires the near-equilibrium condition $|\Pi|\ll\rho$, which clashes with the condition $|\Pi|/\rho>1/3$ needed to accelerate dust \cite{Maartens:1995,Brevik:2017}. While the formal map~\eqref{eq:map} might suggest $|\Qd|/3H\rho\ll1$ as an analogous diffusive near-equilibrium condition, this is not the case: in UG the matter sector remains an exact perfect fluid, and the linear-response law~\eqref{eq:Newton} describes heat exchange between two subsystems in local equilibrium without requiring $\Qd/3H\rho$ to be small. Instead, the ratio $\Qd/3H\rho$ controls the adiabatic sound speed of the unimodular effective fluid, $c_s^2\equiv\dot p_{\rm eff}/\dot\rho_{\rm eff}$ \cite{Cruz:2026logo}. Combining $\dot\rho_{\rm eff}=-3(1+\omega)H\rho$ from Eq.~\eqref{eq:effcons} with $\dot p_{\rm eff}=\omega\dot\rho-\Qd=-3\omega(1+\omega)H\rho-(1+\omega)\Qd$ from Eq.~\eqref{eq:UG2} gives
\begin{equation}
c_s^2=\omega+\frac{\Qd}{3H\rho}=\tilde\omega,\qquad \frac{\Qd}{3H\rho}=c_s^2-\omega .
\label{eq:cs2}
\end{equation}
Since any single effective fluid obeying $3H^2=\rhoeff$ and $\dot H=-\tfrac12(\rhoeff+\peff)$ satisfies the kinematic identity $c_s^2=-1-\ddot H/(3H\dot H)$, Eq.~\eqref{eq:kinematics} yields
\begin{equation}
c_s^2=\frac{j-1}{3(1+q)},
\label{eq:cskin}
\end{equation}
which upon substitution into Eq.~\eqref{eq:cs2} with $\rho=2(1+q)H^2/(1+\omega)$ immediately recovers Eq.~\eqref{eq:Qdj}. Thus, the map $\bar\omega\leftrightarrow\tilde\omega$ connects the total parameter of state of the viscous universe (controlling $\ddot a$) to the effective sound speed of the unimodular universe (controlling perturbations).

Consequently, $\Qd/3H\rho$ is constrained not by cosmic acceleration—which in UG requires $Q>\tfrac12(1+3\omega)\rho$ regardless of $\Qd$—but by perturbative stability and causality of the effective fluid, $0\le c_s^2\le1$ (assuming adiabatic effective perturbations $\delta\peff=c_s^2\,\delta\rhoeff$ \cite{Cruz:2026logo}):
\begin{equation}
-3\omega H\rho\le\Qd\le3(1-\omega)H\rho,
\label{eq:window}
\end{equation}
or $0\le\Qd\le3H\rho$ ($1\le j\le4+3q$) for $\omega=0$. The formal diffusive image of the viscous quintessence window ($\Qd<0$ with $1/3<|\Qd|/3H\rho<1$) is mapped to $-1<c_s^2<-1/3$, where adiabatic perturbations are gradient-unstable. In the finite-bath regime (C), Eqs.~\eqref{eq:Newton} and \eqref{eq:cs2} show that $c_s^2>\omega$ if and only if $T_m>T_D$: heat flowing out of matter into a colder bath stiffens the effective fluid, while heat flowing in softens it. For dust ($\omega=0$), perturbative stability ($c_s^2\ge0$) requires $\Qd\ge0$ ($T_D\le T_m$), whereas the matter limit (A) would restrict the stable regime to $\Qd=0$.

%=====================================================================
\section{Rest mass, temperature and the thermal reservoir}
\label{sec:positivity}
%=====================================================================
\subsection{Dark matter as a gas with rest mass}
\label{sec:gas}

We now consider dark matter as a particle with thermal energy. We describe the matter sector as an ideal gas of particles of mass $m$ and number density $n$,
\begin{equation}
\rho=mn+\varepsilon,\qquad p=nT_m=\omega_{\mathrm{th}}\,\varepsilon,
\label{eq:gas}
\end{equation}
where $\varepsilon$ is the thermal (kinetic) energy density characterized by the kinetic temperature $T_{m}$ and we must have $T_{m}/m \ll 1$ in order to preserve the {\it coldness} of cold dark matter \cite{ArmendarizPicon:2014} and $\omega_{\mathrm{th}}$ is the parameter of state of the thermal component alone, $p=\omega_{\mathrm{th}}\varepsilon$, rather than the total parameter of state $\omega=p/\rho$ of Secs.~\ref{sec:equations}--\ref{sec:cosmography}. Both parameters of state coincide when the rest mass is negligible, $\varepsilon\gg mn$, as for an ultrarelativistic gas with $\omega_{\mathrm{th}}=\omega=1/3$. For cold matter, they differ: $p/\rho\simeq0$, whereas $\omega_{\mathrm{th}}=2/3$ for a non-relativistic monatomic gas, the value used in Fig.~\ref{fig:reservoir}. Keeping the particle number conserved, $\dot n+3Hn=0$, and the particle mass fixed, as the diffusion current allows \cite{Cruz:2026logo,Cardenas:2026}, the rest-mass density obeys $d(mn)/dt+3H\,mn=0$ ($mn\propto a^{-3}$) separately. Substituting Eq.~\eqref{eq:gas} into the balance law~\eqref{eq:UG2} cancels the rest-mass contribution, so that diffusion acts exclusively on the thermal energy:
\begin{equation}
\dot\varepsilon+3(1+\omega_{\mathrm{th}})H\varepsilon=-\Qd ,
\label{eq:epsbalance}
\end{equation}
or equivalently,
\begin{equation}
\frac{d}{dt}\left[\varepsilon\,a^{3(1+\omega_{\mathrm{th}})}\right]=-a^{3(1+\omega_{\mathrm{th}})}\,\Qd .
\label{eq:epsa}
\end{equation}

\emph{Temperature.} Writing $p$ and $T_m$ in terms of $(\rho,n)$ via $\varepsilon=\rho-mn$,
\begin{equation}
p=\omega_{\mathrm{th}}\,(\rho-mn),\qquad T_m=\frac{p}{n}=\omega_{\mathrm{th}}\left(\frac{\rho}{n}-m\right),
\label{eq:pTrhon}
\end{equation}
yields the partial derivatives
\begin{gather}
\left(\frac{\partial p}{\partial\rho}\right)_{\!n}=\omega_{\mathrm{th}},\qquad
\left(\frac{\partial T_m}{\partial n}\right)_{\!\rho}=-\frac{\omega_{\mathrm{th}}\rho}{n^2},\nonumber\\
\left(\frac{\partial T_m}{\partial\rho}\right)_{\!n}=\frac{\omega_{\mathrm{th}}}{n}=\frac{T_m}{\varepsilon},
\label{eq:partials}
\end{gather}
which satisfy the integrability condition~\eqref{eq:integrability}. Inserting them into Eq.~\eqref{eq:Tdot} with $\mathcal F=\Qd$ gives
\begin{equation}
\frac{\dot T_m}{T_m}=-3\omega_{\mathrm{th}} H-\frac{\Qd}{\varepsilon}.
\label{eq:Tideal}
\end{equation}
Combining Eq.~\eqref{eq:Tideal} with Eq.~\eqref{eq:epsbalance} (or directly using $T_m=\omega_{\mathrm{th}}\varepsilon/n$ with $n=n_0a^{-3}$) integrates to
\begin{equation}
\frac{T_m}{T_{m,0}}=\frac{\varepsilon}{\varepsilon_0}\,a^3 ,
\label{eq:Tint}
\end{equation}
confirming that $T_m$ is proportional to the thermal energy per particle and vanishes if and only if $\varepsilon$ does.

In terms of $\varepsilon$, the Newton law~\eqref{eq:Newton} takes the form
\begin{equation}
\Qd=\Gamma\,(\varepsilon-\varepsilon_D),\qquad \Gamma\equiv\frac{\omega_{\mathrm{th}} K}{n}\ge0,\qquad
\varepsilon_D\equiv\frac{nT_D}{\omega_{\mathrm{th}}},
\label{eq:Gamma}
\end{equation}
where $\varepsilon_D$ is the thermal energy density matter would have at the bath temperature and $\Gamma=K/c_V$ is the thermal exchange rate [with isochoric heat capacity $c_V=(\partial\varepsilon/\partial T_m)_n=n/\omega_{\mathrm{th}}$]. Substituting Eq.~\eqref{eq:Gamma} into Eq.~\eqref{eq:Tideal} yields Newton's law of cooling superposed on the adiabatic expansion cooling $T_m\propto a^{-3\omega_{\mathrm{th}}}$,
\begin{equation}
\dot T_m+3\omega_{\mathrm{th}} HT_m=-\Gamma\,(T_m-T_D).
\label{eq:cooling}
\end{equation}
The cold-bath limit $\varepsilon_D=0$ encompasses the microscopic law of Ref.~\cite{Pellecchia:2026}: just as Eq.~\eqref{eq:PPR} drains the kinetic energy $\rho+p=\dot\phi^2$ of a scalar field, its gas analogue $\Qd\propto a^{-3}\varepsilon$ corresponds to Eq.~\eqref{eq:Gamma} with $\varepsilon_D=0$ and $\Gamma\propto a^{-3}$.

\emph{Sound speed and cosmography.} For the gas~\eqref{eq:gas} with $\peff=\omega_{\mathrm{th}}\varepsilon-Q$, evaluating $c_s^2=\dot p_{\rm eff}/\dot\rho_{\rm eff}$ via Eqs.~\eqref{eq:effcons} and \eqref{eq:epsbalance} modifies Eq.~\eqref{eq:cs2} to
\begin{equation}
c_s^2=\frac{1+\omega_{\mathrm{th}}}{\rho+p}\left(\omega_{\mathrm{th}}\varepsilon+\frac{\Qd}{3H}\right),
\label{eq:csgas}
\end{equation}
shifting the adiabatic gas sound speed $\omega_{\mathrm{th}}(1+\omega_{\mathrm{th}})\varepsilon/(\rho+p)$ by $(1+\omega_{\mathrm{th}})\Qd/[3H(\rho+p)]$. Equating Eq.~\eqref{eq:csgas} to the kinematic identity~\eqref{eq:cskin} and using $\rho+p=-2\dot H=2(1+q)H^2$ yields the generalized cosmographic relation
\begin{equation}
\Qd=\frac{2H^3(j-1)}{1+\omega_{\mathrm{th}}}-3\omega_{\mathrm{th}} H\varepsilon ,
\label{eq:Qdjgas}
\end{equation}
which reduces to Eq.~\eqref{eq:Qdj} when $m=0$. In the opposite limit of strictly cold dust ($\varepsilon\equiv0$), conserved particle number and fixed mass enforce $\Qd=0$ and $j=1$: pressureless matter can feed a dynamical $Q$ ($j\neq1$ in Eq.~\eqref{eq:dust}) only if rest-mass energy is converted.

\subsection{Diffusion drains heat, not rest mass}

Inserting the linear closure~\eqref{eq:Gamma} into Eq.~\eqref{eq:epsbalance} gives the first-order linear differential equation
\begin{equation}
\dot\varepsilon+\left[3(1+\omega_{\mathrm{th}})H+\Gamma\right]\varepsilon=\Gamma\varepsilon_D\ \ge0 ,
\label{eq:epsGamma}
\end{equation}
since $\Gamma\ge0$ and $\varepsilon_D\ge0$. Applying the differential form of Gr\"onwall's inequality \cite{Gronwall:1919} with the integrating factor $\exp\{\int_{t_0}^{t}[3(1+\omega_{\mathrm{th}})H+\Gamma]dt'\}$ immediately yields
\begin{equation}
\varepsilon(t)\ \ge\ \varepsilon(t_0)\exp\left\{-\int_{t_0}^{t}\left[3(1+\omega_{\mathrm{th}})H+\Gamma\right]dt'\right\},
\label{eq:positivity}
\end{equation}
where the equality holds for a cold bath ($\varepsilon_D=0$). Because the exponential never vanishes for finite $H$ and $\Gamma$, any state with $\varepsilon(t_0)>0$ maintains $\varepsilon(t)>0$—and thus $T_m>0$ and $\rho>mn$—at all subsequent times: \emph{diffusion compatible with the second law drains heat, never rest mass}. Indeed, for $T_D>0$ this positivity holds for any constitutive law obeying Eq.~\eqref{eq:secondlawC}, since $T_m\propto\varepsilon a^3$ drops below $T_D$ before $\varepsilon$ can reach zero, reversing the sign of $\Qd$. Strict positivity of $\rho+p=mn+(1+\omega_{\mathrm{th}})\varepsilon>0$ has three direct consequences: $\dot H=-\tfrac12(\rho+p)<0$ by Eq.~\eqref{eq:UG1}; the total parameter of state satisfies $1+\wtot=(\rho+p)/(\rho+Q)>0$; and the Bekenstein--Hawking entropy of the apparent horizon $S_h=8\pi^2/H^2$ increases monotonically, $\dot S_h=8\pi^2(\rho+p)/H^3>0$. Thermodynamically consistent diffusion therefore precludes effective phantom behavior at the local level, whereas in GR a viscous fluid can cross into the phantom regime $\Pi<-(\rho+p)$ while respecting the local second law $\Pi\le0$, being ruled out only by the generalized second law at the horizon (Sec.~\ref{sec:horizon}).

\subsection{The thermal reservoir}
\label{sec:reservoir}

Equation~\eqref{eq:epsa} also bounds the total energy that diffusion can deliver to the geometric
sector. Defining the dilution-weighted heat transfer from an initial time $t_1$ to $t\ge t_1$ and using $\varepsilon(t)\ge0$ from Eq.~\eqref{eq:positivity}, we have
\begin{align}
Y(t)&\equiv\int_{t_1}^{t}a^{3(1+\omega_{\mathrm{th}})}\,dQ
=\varepsilon_1a_1^{3(1+\omega_{\mathrm{th}})}-\varepsilon(t)\,a^{3(1+\omega_{\mathrm{th}})}\nonumber\\
&\le\ \varepsilon_1a_1^{3(1+\omega_{\mathrm{th}})}.
\label{eq:Y}
\end{align}
Integrating $Q(t)-Q(t_1)=\int_{t_1}^{t}a^{-3(1+\omega_{\mathrm{th}})}\,dY$ by parts gives
\begin{equation}
Q(t)-Q(t_1)=\frac{Y(t)}{a^{3(1+\omega_{\mathrm{th}})}}+3(1+\omega_{\mathrm{th}})\int_{t_1}^{t}\frac{H\,Y}{a^{3(1+\omega_{\mathrm{th}}})}\,dt' ,
\label{eq:parts}
\end{equation}
where both terms carry non-negative weights in an expanding universe. Inserting the upper bound~\eqref{eq:Y} into Eq.~\eqref{eq:parts} and evaluating the remaining exact integral yields the universal bound
\begin{equation}
Q(t)-Q(t_1)\ \le\ \varepsilon(t_1)\qquad\text{for all } t\ge t_1,
\label{eq:reservoir}
\end{equation}
independent of the intermediate sign of $\Qd$. Because rest mass is unavailable, the growth of the cosmological term cannot exceed the initial thermal energy density of matter. While unrestrictive for relativistic matter ($\varepsilon\simeq\rho$), Eq.~\eqref{eq:reservoir} severely limits cold dark matter ($\omega_{\mathrm{th}}=2/3$), where equipartition gives $\varepsilon/mn=3T_m/2m=\langle v^2\rangle/2\lesssim10^{-6}$ even in massive halos and much less in the homogeneous background. From today onward, thermal diffusion from cold dark matter can increase $Q$ by at most $\Delta Q\lesssim10^{-6}\rho_0$, keeping the expansion indistinguishable from $\Lambda$CDM ($j=1$) to that accuracy. For instance, integrating the cold-bath law ($\varepsilon_D=0$) with constant $\Gamma/3H$ from $a_0=1$ gives $\varepsilon=\varepsilon_0\,a^{-3(1+\omega_{\mathrm{th}})-\Gamma/H}$ and
\begin{equation}
\Delta Q=\frac{\Gamma\,\varepsilon_0}{\Gamma+3(1+\omega_{\mathrm{th}})H}\left[1-a^{-3(1+\omega_{\mathrm{th}})-\Gamma/H}\right]<\varepsilon_0,
\label{eq:DQNewton}
\end{equation}
which respects the reservoir bound for any exchange rate [Fig.~\ref{fig:reservoir}(b)]. Generating a dynamically relevant $Q$ therefore requires either primordial diffusion while matter is hot \cite{Pellecchia:2026} or relaxing particle-number or particle-mass conservation to convert rest mass (Sec.~\ref{sec:discussion}).

\subsection{The logotropic dark fluid}
\label{sec:logo}

\begin{figure}[t]
\centering
\includegraphics[width=\columnwidth]{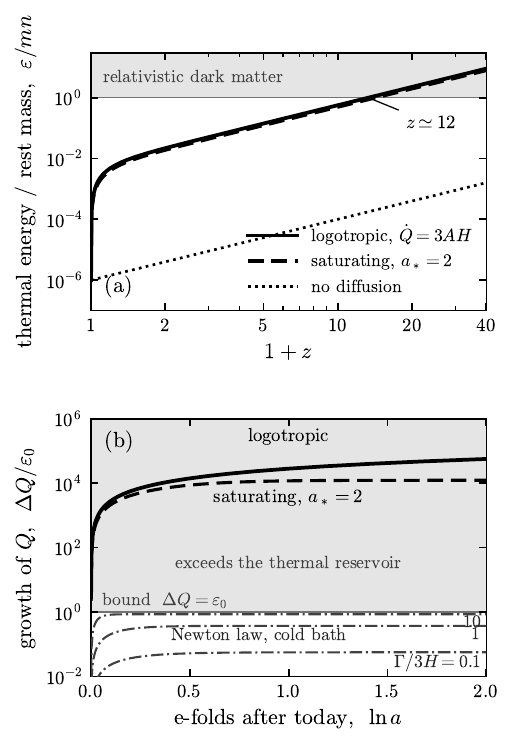}
\caption{Thermal reservoir of dark matter with conserved rest mass, for $\omega_{\mathrm{th}}=2/3$, the values
of Refs.~\cite{Chavanis:2016,Cruz:2026logo}, and $\varepsilon_0/mn_0=10^{-6}$, an upper value for
cold dark matter. (a) Thermal energy per unit rest mass that dark matter must have had in the past
for the energy drained by the logotropic and saturating laws to be heat, Eq.~\eqref{eq:epspast},
compared with adiabatic evolution without diffusion; the shaded region is relativistic. (b) Growth
of the cosmological term after today, in units of the present thermal energy density. The shaded
region violates the bound~\eqref{eq:reservoir}; the cold-bath Newton law,
Eq.~\eqref{eq:DQNewton}, respects it for any exchange rate.}
\label{fig:reservoir}
\end{figure}

The logotropic dark fluid \cite{Chavanis:2016} is equivalent at the background level to dissipative UG with $p\equiv0$ and $Q=3A\ln a+Q_c$ ($\Qd=3AH$), yielding $\rho=\rho_{r,0}a^{-3}-A$ \cite{Cruz:2026logo}, where in units of $3H_0^2$ one has $A=2.56\times10^{-3}$, $\rho_{r,0}=0.274$, and $\rho_0=\rho_{r,0}-A\simeq0.271$. If particle number and mass are conserved, the logotropic drain $\Qd=3AH$ must be supplied by the thermal reservoir via Eq.~\eqref{eq:epsa}, $d[\varepsilon a^{3(1+\omega_{\mathrm{th}})}]/d\ln a=-3A\,a^{3(1+\omega_{\mathrm{th}})}$, which integrates from today to
\begin{equation}
\varepsilon\,a^{3(1+\omega_{\mathrm{th}})}=\varepsilon_0+\frac{A}{1+\omega_{\mathrm{th}}}\left[1-a^{3(1+\omega_{\mathrm{th}})}\right],
\label{eq:epslogo}
\end{equation}
and by Eq.~\eqref{eq:Tint} gives the temperature evolution
\begin{equation}
\frac{T_m}{T_{m,0}}=a^{-3\omega_{\mathrm{th}}}\left[1-\frac{A}{(1+\omega_{\mathrm{th}})\,\varepsilon_0}
\left(a^{3(1+\omega_{\mathrm{th}})}-1\right)\right].
\label{eq:Tlogo}
\end{equation}
Because the logotropic law extracts a fixed energy $3A$ per e-fold regardless of the remaining heat content, and $A/\varepsilon_0\simeq10^4$ for cold dark matter ($\varepsilon_0/mn_0=10^{-6}$), this evolution fails in both time directions:\\
\emph{Forward in time}, $T_m$ and $\varepsilon$ vanish at the scale factor
\begin{equation}
\ln a_T=\frac{\ln\left[1+(1+\omega_{\mathrm{th}})\varepsilon_0/A\right]}{3(1+\omega_{\mathrm{th}})}\simeq\frac{\varepsilon_0}{3A}
\simeq3.5\times10^{-5},
\label{eq:aT}
\end{equation}
beyond which the law extracts heat from matter with $\varepsilon<0$ and $T_m<0$, violating the second law~\eqref{eq:secondlawC} for any $T_D\ge0$ and exceeding the bound~\eqref{eq:reservoir} by orders of magnitude within a fraction of an e-fold [Fig.~\ref{fig:reservoir}(b)]—long before the total density changes sign at $a_M=(\rho_{r,0}/A)^{1/3}$ \cite{Cruz:2026logo}.

\emph{Backward in time}, dividing Eq.~\eqref{eq:epslogo} by $mn\,a^{3(1+\omega_{\mathrm{th}})}$ with $mn=mn_0a^{-3}\simeq\rho_0a^{-3}$ requires the past thermal-to-rest-mass ratio to satisfy
\begin{align}
\frac{\varepsilon}{mn}&=\frac{\varepsilon_0}{mn_0}\,a^{-3\omega_{\mathrm{th}}}
+\frac{A}{(1+\omega_{\mathrm{th}})\,mn_0}\left(a^{-3\omega_{\mathrm{th}}}-a^{3}\right)\nonumber\\
&\ge\ \frac{A}{(1+\omega_{\mathrm{th}})\rho_0}\left(a^{-3\omega_{\mathrm{th}}}-a^{3}\right),
\label{eq:epspast}
\end{align}
which for $\omega_{\mathrm{th}}=2/3$ reaches the relativistic regime ($\varepsilon\simeq mn$) at $z\simeq12$ [Fig.~\ref{fig:reservoir}(a)], incompatible with cold dark matter during structure formation. (Identifying $\rho_r$ directly with $mn$ would instead imply $\varepsilon=\rho-mn=-A<0$ at all times.) Even the saturating law $\Qd=3AH/[1+(a/a_*)^3]^2$ of Ref.~\cite{Cruz:2026logo}, which keeps $\rho>0$ for $a_*<a_M$, transfers after today a total energy
\begin{align}
\Delta Q&=A\int_{a_*^{-3}}^{\infty}\frac{dx}{x(1+x)^2}\nonumber\\
&=A\left[\ln\left(1+a_*^3\right)-\frac{a_*^3}{1+a_*^3}\right],
\label{eq:DQsat}
\end{align}
giving $\Delta Q=1.31A\simeq10^{4}\varepsilon_0$ for $a_*=2$. Thus, both the logotropic and saturating laws require rest-mass conversion rather than thermal conduction.

\begin{table}[t]
\begin{ruledtabular}
\begin{tabular}{lll}
Law & Removes & Status in regime (C)\\
\hline
$Q={\rm const}$ ($\Lambda$CDM) & nothing & consistent\\
$\Qd=\Gamma\varepsilon$ (cold bath) & heat & $\Delta Q\le\varepsilon_0$\\
$\Qd=\xi\rho$, $\xi>0$ & rest mass & needs conversion\\
$Q=\alpha\rho$, $\alpha>0$ & none (heats) & while $T_m<T_D$\\
$\Qd=3AH$ (logotropic) & rest mass & fails at $a_T\simeq1$\\
Saturating & rest mass & $\Delta Q\simeq1.3A$\\
\end{tabular}
\end{ruledtabular}
\caption{Diffusion laws for dark matter with conserved rest mass, read through
Eq.~\eqref{eq:Gamma}. The second column states what each law removes from matter.}
\label{tab:laws}
\end{table}

Table~\ref{tab:laws} summarizes the thermodynamic status of representative diffusion laws in the finite-bath regime (C). Density-proportional laws $\Qd=\xi\rho$ ($\xi>0$) \cite{Josset:2017,Corral:2020,Cruz:2024} continue to drain energy at $\varepsilon=0$ ($\Qd=\xi mn>0$) and thus require rest-mass conversion, whereas the barotropic law $Q=\alpha\rho$ ($\alpha>0$) gives $\Qd=-3\alpha H(\rho+p)/(1+\alpha)<0$, heating matter and remaining admissible only while $T_m<T_D$.

%=====================================================================
\section{The apparent horizon}
\label{sec:horizon}

The apparent horizon of a spatially flat FLRW universe is located at $r_A=1/H$, with Bekenstein--Hawking entropy $S_h=8\pi^2/H^2$, enclosed volume $V_h=4\pi/3H^3$, and dynamical Hayward--Kodama temperature
\begin{equation}
T_h=\frac{H}{2\pi}\left(1+\frac{\dot H}{2H^2}\right)=\frac{H}{4\pi}(1-q),
\label{eq:Thq}
\end{equation}
which is positive for $q<1$ \cite{Bak:2000,Cai:2005,Wang:2006}; by Eq.~\eqref{eq:kinematics}, $\dot V_h/V_h=-3\dot H/H=3(1+q)H$. To evaluate the horizon entropy budget, we adopt the standard local horizon-equilibrium prescription $T_m=T_h$ for the fluid energy crossing the horizon (distinct from the microscopic kinetic temperature of Sec.~\ref{sec:positivity}), so that $T_h\dot S_{\rm in}=V_h[\dot\rho+(\rho+p)\dot V_h/V_h]$, and in the finite-bath regime (C) we include the bath entropy $\dot S_{D,{\rm in}}=V_h\Qd/T_D$ inside the horizon, consistent with its vanishing energy flux $\rho_Q+p_Q=0$. In UG, Eq.~\eqref{eq:UG2} gives $\dot\rho+(\rho+p)\dot V_h/V_h=3qH(\rho+p)-\Qd$. Using $\rho+p=2H^2(1+q)$, $\dot S_h=-16\pi^2\dot H/H^3=16\pi^2(1+q)/H$, and Eq.~\eqref{eq:Thq}, the total entropy rate combines into
\begin{align}
\dot S_h+\dot S_{\rm in}+\dot S_{D,{\rm in}}&=\frac{16\pi^2(1+q)^2}{H(1-q)}\nonumber\\
&\quad+V_h\Qd\left(\frac{1}{T_D}-\frac{1}{T_h}\right).
\label{eq:GSLUG}
\end{align}
In GR, Eq.~\eqref{eq:GR2} gives instead $\dot\rho+(\rho+p)\dot V_h/V_h=3qH(\rho+p)-3H\Pi$; using $(1+\omega)\rho=2H^2(1+q)-\Pi$ and $3H^2=\rho$ from Eq.~\eqref{eq:GR1} together with Eq.~\eqref{eq:Thq}, we obtain
\begin{align}
\dot S_h+\dot S_{\rm in}&=\frac{16\pi^2(1+q)(2-q+3\omega)\rho}{3H^3(1-q)}\nonumber\\
&=\frac{16\pi^2(1+q)}{H(1-q)}\left(1+q-\frac{\Pi}{H^2}\right).
\label{eq:GSLGR}
\end{align}
For $q<1$, the first term of Eq.~\eqref{eq:GSLUG} is manifestly non-negative and the second is non-negative by the local second law~\eqref{eq:secondlawC}, so the generalized second law (GSL) in UG follows directly from the local one. (In the reversible limit (B), only the first term survives, rendering the GSL blind to diffusion even on a phantom branch, which only $\dot S_h$ detects \cite{Cruz:2026logo}.) In GR, because $\omega>-1$ ensures $2-q+3\omega=(1-q)+(1+3\omega)>0$ across the accelerating and phantom regimes ($q<0$), Eq.~\eqref{eq:GSLGR} makes the GSL for $q<1$ equivalent to $q\ge-1$, i.e., $\Pi\ge-(1+\omega)\rho$. Thus, both frameworks exclude effective phantom behavior via thermodynamics, but by different routes: in GR the horizon is required to rule out the viscous phantom regime allowed by $\Pi\le0$, whereas in UG the local second law and temperature positivity suffice.

%=====================================================================
\section{Discussion}
\label{sec:discussion}

\emph{Friction, warm rather than cold.} A natural image for unimodular dissipation is
inflationary friction, $3H\dot\phi\simeq-V'(\phi)$. Hubble friction, however, is reversible and produces no entropy. The dissipative counterpart is warm inflation \cite{Berera:1995}, where an additional friction $\Upsilon\dot\phi$ feeds a radiation bath and comes with thermal noise. The Langevin equation of Ref.~\cite{Pellecchia:2026} has the same structure, with $\bar\beta m_p$ in place of $\Upsilon$; the difference is that the warm-inflation bath is visible radiation, whereas the unimodular bath is hidden and gravitates only through $\Lambda$. The drained energy is what accelerates the expansion, independently of the integration constant $\Lambda_0$, which can vanish; identifying $Q$ with a running vacuum would obscure this.

\emph{Apparent phantom crossing.} The observational debate about a phantom crossing concerns the dark-energy parameter of state reconstructed under the assumption of conserved matter \cite{DESI:2025}, not $\wtot$. If unimodular data are analyzed as GR with dust
$\rho_{\rm fid}\propto a^{-3}$ plus a dark-energy component, then
$1+w_{\rm DE}^{\rm eff}=(\rho-\rho_{\rm fid})/\rho_{\rm DE}^{\rm eff}$, which changes sign where $\rho a^3$ crosses its fiducial value, without any violation of the null energy condition. With a cold bath, $\rho a^3$ decreases, so the reconstructed parameter of state would go from quintessence-like to phantom-like as time proceeds, opposite to the trend reported in Ref.~\cite{DESI:2025}; moreover, by Eq.~\eqref{eq:reservoir}, cold dark matter can lower $\rho a^3$ only by a fraction of order $\varepsilon/\rho$. Reproducing the reported trend through diffusion of dark matter requires $\Qd<0$, a bath hotter than the dark matter, which heats it to $\varepsilon/\rho$ of the order of the required change in $\rho a^3$ and, by Eq.~\eqref{eq:csgas}, lowers the effective sound speed. This chain deserves a dedicated analysis, along the lines of the interacting-vacuum reading of unimodular diffusion \cite{Kashyap:2026}.

\emph{Open problems.} Four points are left open.
(i) The bath temperature $T_D$ and $K$ are microscopic inputs; in the model of
Ref.~\cite{Pellecchia:2026} they descend from the primordial state of the hidden degrees of freedom and from the friction coefficient $\bar\beta$, and whether $T_D$ evolves depends on the heat capacity of the bath. (ii) Perturbations require the diffusion current $J_\mu$ beyond FLRW, which neither the phenomenological nor the microscopic description fixes; our statements on $c_s^2$ hold for adiabatic perturbations of the effective fluid. (iii) A dynamically relevant $Q$ fed by cold dark matter requires rest-mass conversion (Sec.~\ref{sec:reservoir}). Particle annihilation into the bath, $\dot n+3Hn=n\Gamma_c$ with $\Gamma_c<0$, achieves it within the open-system formalism of Refs.~\cite{Prigogine:1989,Cardenas:2026}, where $nT\dot\sigma=-3Hp_c-(\rho+p)\Gamma_c-\Qd$, with $p_c$ the creation pressure, and the bath term of Eq.~\eqref{eq:Stot} can be added without modification. A decrease of the particle mass at fixed particle number, as for mass-varying particles, transfers energy as work rather than heat: it leaves the matter entropy unchanged, lies outside Eq.~\eqref{eq:Stot}, and would make the logotropic correspondence adiabatic, as asserted in Ref.~\cite{Cruz:2026logo}, for a reason different from the reversible limit (B). (iv) The Newton law~\eqref{eq:Newton} is an instantaneous closure. A causal generalization, in which the exchange flux relaxes toward $K(T_m-T_D)$ over a finite time, would be the analogue of the Israel--Stewart extension of Eckart's theory; it is suggested by baths whose damping kernel is non-local in time \cite{Pellecchia:2026}. It would require revisiting Eq.~\eqref{eq:positivity}: with thermal inertia, $\Qd\,(T_m-T_D)$ can become transiently negative, and the positivity of $\varepsilon$ would then depend on the relaxation time.

%=====================================================================
\section{Concluding remarks}
\label{sec:conclusions}

We have shown that the formal correspondence $\Pi\leftrightarrow\Qd/3H$ between bulk viscosity in GR and energy diffusion in UG is restricted to the internal balance, entropy, and temperature equations of the matter sector. Once the Friedmann constraint and the entropy of the receiving sector are included, the two mechanisms describe fundamentally different thermodynamic processes. Assigning a temperature $T_D$ to the hidden reservoir that absorbs the diffused energy unifies the matter ($T_D\to\infty$) and adiabatic ($T_D\to T_m$) prescriptions in the literature as limiting cases of the open-system entropy production~\eqref{eq:Stot}, while the microscopic fluctuation-dissipation framework of Ref.~\cite{Pellecchia:2026} selects a finite-temperature bath.

Under this open-system accounting (summarized in Table~\ref{tab:dictionary}), diffusion operates as heat conduction governed by $\Qd=K(T_m-T_D)$ rather than internal expansion work. While bulk viscosity can only heat a closed fluid and is determined cosmographically by $q$ (rendering $\Lambda$CDM permanently dissipative), diffusion into a cold bath drains heat from an open fluid, is determined by the jerk $j$ (making $\Lambda$CDM its adiabatic limit), and shifts the effective sound speed by $c_s^2-\omega=\Qd/3H\rho$ independently of cosmic acceleration. Furthermore, when dark matter is modeled as a gas of massive particles with conserved rest mass, the second law prevents diffusion from extracting rest-mass energy, ensuring $T_m>0$, ruling out phantom behavior locally, and guaranteeing the generalized second law at the dynamical apparent horizon. Because the growth of the cosmological term is bounded by the available thermal energy, $\Delta Q\le\varepsilon(t_1)$, cold dark matter with conserved particle number and mass cannot source a late-time logotropic dark fluid without having been relativistic at $z\simeq12$.

The natural continuation is a microscopic determination of $T_D$ and $K$, the treatment of
rest-mass conversion into the bath, and the extension of the diffusion current to perturbations, which together would turn the thermal reading of unimodular diffusion into an observational one.

\begin{acknowledgments}
MC work was partially supported by S.N.I.I. (SECIHTI-M\'exico). SL acknowledges the FONDECYT
grant N$^{\circ}$ 1250969, Chile.
\end{acknowledgments}

%=====================================================================
\appendix
\section{Temperature evolution}
\label{app:temperature}

For matter with conserved particle number, the Gibbs relation per particle reads
$d\sigma=d\rho/(nT)-(\rho+p)\,dn/(n^2T)$. Exactness of $d\sigma$ in the variables $(\rho,n)$
requires the equality of mixed derivatives,
$\partial_n[1/(nT)]=-\partial_\rho[(\rho+p)/(n^2T)]$. Carrying out the derivatives and multiplying
by $-n^2T^2$ gives
\begin{equation}
n\frac{\partial T}{\partial n}+(\rho+p)\frac{\partial T}{\partial\rho}=T\frac{\partial p}{\partial\rho},
\label{eq:integrability}
\end{equation}
a kinematical identity independent of the dynamics \cite{Maartens:1996}. Along the flow, with
$\dot n=-3Hn$, $\dot T=(\partial T/\partial\rho)_n\dot\rho-3Hn(\partial T/\partial n)_\rho$.
Replacing $n\,\partial T/\partial n$ by means of Eq.~\eqref{eq:integrability},
\begin{equation}
\frac{\dot T}{T}=-3H\left(\frac{\partial p}{\partial\rho}\right)_{\!n}
+\frac{1}{T}\left(\frac{\partial T}{\partial\rho}\right)_{\!n}\left[\dot\rho+3H(\rho+p)\right],
\end{equation}
which with Eqs.~\eqref{eq:GR2} and \eqref{eq:UG2} gives Eq.~\eqref{eq:Tdot}.

For $p=\omega\rho$ the general solution of Eq.~\eqref{eq:integrability} is
\begin{equation}
T=\rho^{\omega/(1+\omega)}F(y),\qquad y\equiv\frac{\rho^{1/(1+\omega)}}{n},
\end{equation}
with $F$ arbitrary; the ideal gas without rest mass is $F(y)=\omega y$. The gas with rest mass of
Sec.~\ref{sec:positivity}, Eq.~\eqref{eq:pTrhon}, is not of the form $p=\omega\rho$, but it also
satisfies Eq.~\eqref{eq:integrability}, as shown below Eq.~\eqref{eq:partials}, and reduces to
$F(y)=\omega_{\mathrm{th}} y$ for $m=0$. Along the flow,
\begin{equation}
\frac{\dot y}{y}=\frac{\dot\rho+3H(1+\omega)\rho}{(1+\omega)\rho}=
\begin{cases}
-\dfrac{3H\Pi}{(1+\omega)\rho} & \text{(GR)},\\[2ex]
-\dfrac{\Qd}{(1+\omega)\rho} & \text{(UG)},
\end{cases}
\end{equation}
so $y$ is conserved only without dissipation. For the ideal gas this means that the power law
$T\propto\rho^{\omega/(1+\omega)}$ of the adiabatic case fails as soon as $\Pi$ or $\Qd$ is
nonzero: the temperature depends on $\rho$ and $n$ separately. A barotropic temperature (constant
$F$) remains an admissible equation of state, but it is incompatible with $p=nT$ once $y$ evolves.
This is the precise sense in which particle number is conserved while the energy per particle
changes \cite{Cruz:2026logo}, and the reason why we use the ideal-gas closure in
Sec.~\ref{sec:positivity}, which ties $T_m$ to the thermal energy per particle.

%=====================================================================


\begin{thebibliography}{99}

\bibitem{Unruh:1989}
W.~G.~Unruh, Phys.\ Rev.\ D \textbf{40}, 1048 (1989).

\bibitem{Ellis:2011}
G.~F.~R.~Ellis, H.~van Elst, J.~Murugan, and J.-P.~Uzan, Classical Quantum Gravity
\textbf{28}, 225007 (2011).

\bibitem{Carballo-Rubio:2022}
R.~Carballo-Rubio, L.~J.~Garay, and G.~Garc\'ia-Moreno, Classical Quantum Gravity
\textbf{39}, 243001 (2022).

\bibitem{Bengochea:2023}
G.~R.~Bengochea, G.~Le\'on, A.~Perez, and D.~Sudarsky, J.\ Cosmol.\ Astropart.\ Phys.\
11 (2023) 011.

\bibitem{Josset:2017}
T.~Josset, A.~Perez, and D.~Sudarsky, Phys.\ Rev.\ Lett.\ \textbf{118}, 021102 (2017).

\bibitem{Perez:2019}
A.~Perez and D.~Sudarsky, Phys.\ Rev.\ Lett.\ \textbf{122}, 221302 (2019).

\bibitem{Garcia-Aspeitia:2019}
M.~A.~Garc\'ia-Aspeitia, C.~Mart\'inez-Robles, A.~Hern\'andez-Almada, J.~Maga\~na, and
V.~Motta, Phys.\ Rev.\ D \textbf{99}, 123525 (2019).

\bibitem{Corral:2020}
C.~Corral, N.~Cruz, and E.~Gonz\'alez, Phys.\ Rev.\ D \textbf{102}, 023508 (2020).

\bibitem{Perez:2021}
A.~Perez, D.~Sudarsky, and E.~Wilson-Ewing, Gen.\ Relativ.\ Gravit.\ \textbf{53}, 7 (2021).

\bibitem{Linares:2021}
F.~X.~Linares Cede\~no and U.~Nucamendi, Phys.\ Dark Universe \textbf{32}, 100807 (2021).

\bibitem{Landau:2023}
S.~J.~Landau, M.~Benetti, A.~Perez, and D.~Sudarsky, Phys.\ Rev.\ D \textbf{108}, 043524
(2023).

\bibitem{Cruz:2024}
M.~Cruz, N.~Cruz, and S.~Lepe, Eur.\ Phys.\ J.\ C \textbf{84}, 1186 (2024).

\bibitem{Pellecchia:2026}
P.~Pellecchia, A.~Perez, and S.~Ribisi, Phys.\ Rev.\ D \textbf{114}, 046030 (2026),
arXiv:2607.03272.

\bibitem{Eckart:1940}
C.~Eckart, Phys.\ Rev.\ \textbf{58}, 919 (1940).

\bibitem{Israel:1979}
W.~Israel and J.~M.~Stewart, Ann.\ Phys.\ (N.Y.) \textbf{118}, 341 (1979).

\bibitem{Hiscock:1985}
W.~A.~Hiscock and L.~Lindblom, Phys.\ Rev.\ D \textbf{31}, 725 (1985).

\bibitem{Maartens:1995}
R.~Maartens, Classical Quantum Gravity \textbf{12}, 1455 (1995).

\bibitem{Maartens:1996}
R.~Maartens, \emph{Causal thermodynamics in relativity}, arXiv:astro-ph/9609119.

\bibitem{Brevik:2017}
I.~Brevik, \O.~Gr\o n, J.~de~Haro, S.~D.~Odintsov, and E.~N.~Saridakis, Int.\ J.\ Mod.\
Phys.\ D \textbf{26}, 1730024 (2017).

\bibitem{Cruz:2026logo}
M.~Cruz and S.~Lepe, \emph{The logotropic dark fluid as an energy diffusion in unimodular
gravity}, arXiv:2609.21006.

\bibitem{Cardenas:2026}
V.~H.~C\'ardenas, M.~Cruz, and S.~Lepe, Phys.\ Lett.\ B \textbf{881}, 140955 (2026), arXiv:2609.29569.

\bibitem{Aguilar:2026}
G.~Aguilar-P\'erez, M.~Cruz, and S.~Lepe, \emph{Spatial curvature in unimodular gravity},
arXiv:2605.16751.

\bibitem{Chavanis:2016}
P.-H.~Chavanis, Phys.\ Lett.\ B \textbf{758}, 59 (2016).

\bibitem{Caldeira:1983}
A.~O.~Caldeira and A.~J.~Leggett, Physica A \textbf{121}, 587 (1983).

\bibitem{Calzetta:2008}
E.~A.~Calzetta and B.-L.~B.~Hu, \emph{Nonequilibrium Quantum Field Theory} (Cambridge
University Press, Cambridge, England, 2008).

\bibitem{ArmendarizPicon:2014}
C.~Armendariz-Picon and J.~T.~Neelakanta, J.\ Cosmol.\ Astropart.\ Phys.\ 03, 049 (2014), 	arXiv:1309.6971.

\bibitem{Gronwall:1919}
T.~H.~Gronwall, Ann.\ Math.\ \textbf{20}, 292 (1919).

\bibitem{Bak:2000}
D.~Bak and S.-J.~Rey, Classical Quantum Gravity \textbf{17}, L83 (2000).

\bibitem{Cai:2005}
R.-G.~Cai and S.~P.~Kim, J.\ High Energy Phys.\ 02 (2005) 050.

\bibitem{Wang:2006}
B.~Wang, Y.~Gong, and E.~Abdalla, Phys.\ Rev.\ D \textbf{74}, 083520 (2006).

\bibitem{Berera:1995}
A.~Berera, Phys.\ Rev.\ Lett.\ \textbf{75}, 3218 (1995).

\bibitem{DESI:2025}
M.~Abdul-Karim \emph{et al.} (DESI Collaboration), Phys.\ Rev.\ D \textbf{112}, 083515 (2025), arXiv:2503.14738.

\bibitem{Kashyap:2026}
G.~Kashyap and N.~K.~Singh, \emph{Unimodular diffusion and interacting vacuum cosmology},
arXiv:2603.21675.

\bibitem{Prigogine:1989}
I.~Prigogine, J.~Geheniau, E.~Gunzig, and P.~Nardone, Gen.\ Relativ.\ Gravit.\ \textbf{21},
767 (1989).

\end{thebibliography}
\end{document}